\documentclass[preprint,12pt]{elsarticle}

\makeatletter
\def\ps@pprintTitle{%
 \let\@oddhead\@empty
 \let\@evenhead\@empty
 \def\@oddfoot{\centerline{\thepage}}%
 \let\@evenfoot\@oddfoot}
\makeatother 

\usepackage{amssymb}
\usepackage{amsmath}
\usepackage{graphicx}
\usepackage{epstopdf}
\usepackage{natbib}
\usepackage{graphicx}
\usepackage{color}

\newcommand{\vx}{{\mathbf x}}
\newcommand{\va}{{\mathbf a}}

\newcommand{\D}{{\mathbf D}}
\newcommand{\vk}{{\mathbf k}}

\journal{Advances in Water Resources}

\begin{document}

\begin{frontmatter}



\title{Pictures of Blockscale Transport: Effective versus Ensemble Dispersion and Its Uncertainty}


\author[label1]{Felipe P. J. de Barros}
\author[label2]{Marco Dentz}
\address[label1]{Sonny Astani Department of Civil and Environmental Engineering, University of Southern California, Los Angeles, California, USA}
\address[label2]{Institute of Environmental Assessment and Water Research (IDAEA), Spanish National Research Council (CSIC), Barcelona, Spain}


\begin{abstract}
Solute transport models tend to use coarse numerical grid blocks to
alleviate computational costs. Aside from computational issues, the
subsurface environment is usually characterized over a coarse
measurement network where only large scale fluctuations of the flow
field are captured. Neglecting the subscale velocity fluctuations in
transport simulators can lead to erroneous predictions with
consequences in risk analysis and remediation. For such reasons,
upscaled dispersion coefficients in spatially heterogeneous flow
fields must (1) account for the subscale variability that is
filtered out by homogenization and (2) be modeled as a random
function to incorporate the uncertainty associated with non-ergodic
solute bodies. In this work, we examine the low order statistical
properties of the blockscale dispersion tensor. The blockscale is
defined as the scale upon which the spatially variable flow field is
homogenized (e.g. the numerical grid block). Using a Lagrangian
framework, we discuss different conceptualizations of the blockscale
dispersion tensor. We distinguish effective and ensemble blockscale
dispersion, which measure the impact of subscale velocity
fluctuations on solute dispersion. Ensemble dispersion quantifies
subscale velocity fluctuations between realizations, which
overestimates the actual velocity variability. Effective dispersion
on the other hand quantifies the actual blockscale velocity
variability and thus reflects the impact of subscale velocity
fluctuations on mixing and spreading. Based on these concepts, we
quantify the impact of subscale velocity fluctuations on solute
particle spreading and determine the governing equations for the
coarse-grained concentration distributions. We develop analytical
and semi-analytical expressions for the average and variance of the
blockscale dispersion tensor in 3D flow fields as a function of the
structural parameters characterizing the subsurface. Our results
illustrate the relevance of the blockscale, the initial scale of the
solute body and local-scale dispersion in controlling the
uncertainty of the plume's dispersive behavior. The analysis
performed in this work has implications in numerical modeling (i.e.
grid design) and allows to quantify the uncertainty of the
blockscale dispersion tensor.
\end{abstract}



\end{frontmatter}


\section{Introduction}\label{sec:introduction}

Natural porous media are spatially variable and characterized by a
distribution of heterogeneity scales. This multi-scale variability
is an important factor controlling flow and large scale dispersive
behavior of a solute plume. A common approach to model the effects
of heterogeneity in solute transport is through the use of
dispersion coefficients. Defining dispersion coefficients allows to
quantify the impact of the subscale velocity fluctuations on solute
spreading. In general, dispersion coefficients are not constant and
evolve in time \cite{gelhar1983,dagan1988} and in many cases, the
observed large scale transport displays a non-Fickian behavior
\cite{fiori2003p2,LeBorgne2010AWR,dentz2011JCH}. The key question is
in determining the subscale relevant to solute dispersion.

The difficulty associated with transport modeling in natural porous
media is that an accurate reproduction of the spatial structure of
the hydraulic properties at all scales is not achievable within the
context of applications. This is mainly because of technological and
financial budget constraints \cite{rubin2003book}. As a consequence,
the spatial variability of the hydrogeological properties is
characterized over a coarse measurement network. Hence only large
scale features of the hydraulic conductivity fluctuations are
deterministically captured at the field site and this information
will be transferred to the numerical grid of the flow and transport
simulator \cite{dagan1994CMWR, beckie1996,beckie1996theory,
mehrabi1997, rubin1999,attinger2003}. However, when modeling solute
transport, it is important to account for subscale variability since
velocity fluctuations at scales smaller than the solute body tend to
distort the plume and create new concentration gradients. Thus, the
subscale is defined by the resolution scale imposed by measurements
as well as computational resources (i.e. numerical grid). Neglecting
the small scale variability not captured directly on the numerical
grid can potentially lead to erroneous transport predictions and
consequently, wrong decision making when managing contaminated
sites.

To compensate for the subscale variability in solute transport
modeling, Rubin et al. \cite{rubin1999} developed the concept of
block-effective dispersion. The main idea proposed in Rubin et al.
\cite{rubin1999,rubin2003wrr} consists in developing an ensemble
based block-effective dispersion tensor that accounts for the
homogenized blockscale variability without duplicating the large
scale heterogeneity of the conductivity field directly captured on
the numerical grid. This concept allows flexibility in numerical
grid design without neglecting the dispersive flux of the unmodeled
variability and alleviates the computational burden associated with
fine grid resolution numerical simulations
\cite{bellin2004SERRA,eberhard2005upscaling,fernandez2007wrr,lawrence2007,cortinez2013AGU}.
Similar blockscale dispersion analysis was performed through the use
of volume averaging \cite{wang1999serra,wood2003}.

Theoretical expressions for the \textit{ensemble} dispersion tensor
were obtained by fully homogenizing the heterogeneous porous
formation
\cite{gelhar1983,dagan1984,neuman1987,quinodoz1993,hsu1996}. The
ensemble dispersion theory (also known as the macrodispersion
theory) in the aforementioned works is valid under ergodic
conditions, e.g., when the dimensions of solute body perpendicular
to the mean flow direction is much larger the correlation scale of
heterogeneity. However, in many applications, the characteristic
solute plume dimensions are comparable to the scale of heterogeneity
thus leading to non-ergodic transport conditions; this implies that
dispersion coefficients are prone to uncertainty. With the aim of
studying dispersion under non-ergodic conditions, \textit{effective}
dispersion coefficients were developed to predict the expected
dispersion behavior within a single realization of the hydraulic
conductivity field
\cite[e.g.,][]{kitanidis1988,dagan1990,dagan1991,rajaram1993wrr,fiori1998,zhang_difederico1998,dentz2000point,dentz2000ext}.
The \textit{effective} dispersion tensor converges to the
\textit{ensemble} dispersion tensor when the solute plume samples
the full heterogeneous structure of the geological formation.
Furthermore, as discussed in the literature, the effective
dispersion coefficient for a point source quantifies mixing whereas
the ensemble dispersion spreading
\cite{attinger1999,dentz2000ext,bolster2012JFM}.

Therefore, in order to accurately describe the dispersion behavior
of a solute plume in random flow fields, the dispersion tensor must:
(1) be modeled as a random function since the evolution of the
non-ergodic solute body is not deterministic; and (2) must
incorporate the subscale variability of the velocity field that is
lost due to homogenization (such as in a numerical grid). Hence,
dispersion tensor quantifying large-scale transport is a random
function and should be evaluated at the blockscale. To address this
challenge, de Barros and Rubin \cite{deBarros2011jfm} developed a
modeling framework that treats the dispersion tensor as a random
function. The modeling framework proposed by de Barros and Rubin
\cite{deBarros2011jfm} incorporates the scale of the solute body and
the scale of the grid block. Under different limiting cases, de
Barros and Rubin \cite{deBarros2011jfm} showed that the blockscale
dispersion random function model recovers the conditions explored in
the past. The authors \cite{deBarros2011jfm} developed expressions
for the blockscale dispersion variance for a finite size solute body
released in a stratified geological formation under pure advective
transport and, through the minimum relative entropy principle,
inferred a model for the probability density function of the
blockscale dispersion tensor. Within the context of a fully
homogenized aquifer system (e.g., infinite blockscale), Dentz and de
Barros \cite{dentz_deBarros2013} derived expressions for the
dispersion variance tensor in two- and three-dimensional
statistically anisotropic random flow fields in the presence of
local-scale dispersion. The evolution of the dispersion variance
provides information on the mixing efficiency of the transport
processes. As the solute plume becomes better mixed, the dispersion
variance decreases since the variability between realization
reduces. The authors \cite{dentz_deBarros2013} showed that the
self-averaging behavior of the solute plume strongly depends on the
flow dimensionality and the geometrical properties of the injection
source. Self-averaging implies that the dispersion variance tends to
zero for large times (e.g. asymptotic ergodicity). The
self-averaging properties of the fully upscaled dispersion
coefficient was also studied in a two-dimensional stratified random
flow field \cite{clincy2001} and multi-Gaussian random flows
\cite{suciu2006,suciu2014}.

The aim of this work is to develop and discuss a dispersion concept
that reflects the actual blockscale velocity variability and its
impact on solute mixing and spreading. In addition, we quantify the
sample-to-sample fluctuations of the blockscale dispersion through
its variance. We provide a rational framework that allows (1) to
determine dispersion coefficients in coarse-grained models and (2)
quantify the corresponding uncertainty in the blockscale dispersion
tensor. The mean and variance of the dispersion coefficient are
determined as a function of the blockscale, the characteristic size
of the solute plume, statistical anisotropy ratio of the
heterogeneous permeability field and local dispersion. We develop
novel semi-analytical expressions for the first two statistical
moments of the blockscale dispersion in 3D flow fields under finite
P\'{e}clet conditions.
Our results shed new light on how the characteristic length scales
defining the transport problem control the level of uncertainty in
transport predictions. Specifically, we illustrate the potential of
the scale of the homogenizing region in reducing the uncertainty of
dispersion coefficients.

\section{Basics and Methodology}\label{sec:basics}

\subsection{Flow and Transport}\label{subsec:flow_transp}

The physical problem investigated in this work consists of a
$d$-dimensional steady state, spatially heterogeneous,
incompressible Darcy flow field. The fluctuation of the flow field
stems from the heterogeneity of the geological medium's hydraulic
conductivity. The hydraulic conductivity $K\left(\mathbf{x}\right)$
is locally isotropic and spatially variable in $\mathbf{x} =
\left(x_{1},..., x_{d}\right)^T$. Flow is assumed to be
uniform-in-the-mean along the $x_1$ direction. For the purpose of
this work, we consider flow to be divergence free in the absence of
sinks/sources and far from boundary effects. Furthermore, the
effective porosity $\phi$ of the medium is assumed to be constant.
The divergence-free velocity field
$\mathbf{v}\left(\mathbf{x}\right)$ is determined via Darcy's law
and the flow equation
\begin{align}
\mathbf{v}\left(\mathbf{x}\right)=-K\left(\mathbf{x}\right)\nabla
h\left(\mathbf{x}\right)/\phi, && \nabla\cdot\left[K\left(\mathbf{x}\right)\nabla
h\left(\mathbf{x}\right)\right]  =  0.
\label{eq:PDE-H}
\end{align}
where $h\left(\mathbf{x}\right)$ is the hydraulic head.

An inert dissolved solute is instantaneously released at time
$t=t_0$ and the plume is advected and dispersed according to the
governing equation
\begin{align}
\frac{\partial c \left(\mathbf{x},t\right)}{\partial t} +
\mathbf{v}\left(\mathbf{x}\right)\cdot\nabla c
\left(\mathbf{x},t\right) & =  \nabla \cdot [\mathbf{D}\nabla
c\left(\mathbf{x},t\right)] \label{eq:PDE-C}
\end{align}
where $c\left(\mathbf{x},t\right)$ is the resident concentration of
the dissolved passive solute.
The initial solute distribution is $c\left(\mathbf{x},t=t_{0}\right)
= \rho\left(\mathbf{x}\right)$ and the local-scale dispersion tensor
$\mathbf{D}$ is assumed to be diagonal, e.g. $D_{ij} = D_{ii}
\delta_{ij}$ with $\delta_{ij}$ denoting Kronecker's delta. For
simplicity, we assume $t_{0}=0$

Solute transport can be described equivalently in terms
of the equations of motion of solute particles by the Langevin
equation~\cite[][]{Risken:1996}
\begin{align}
\label{langevin}
\frac{d \mathbf
  x(t|\mathbf a)}{d t} = \mathbf v[\mathbf x(t|\mathbf
a)] + \sqrt{2 \mathbf D} \cdot
\boldsymbol \xi(t),
\end{align}
where $\boldsymbol \xi(t)$ denotes a Gaussian white noise with zero
mean and covariance $\langle \xi_i(t) \xi_j(t^\prime) \rangle =
\delta_{ij} \delta(t - t^\prime)$ with $\delta_{ij}$ the Kronecker
delta and $\delta(t)$ corresponding to the Dirac delta. In this
work, $\boldsymbol \xi(t)$  will be termed \textit{local noise}
since it is related to the local scale dispersion tensor
$\mathbf{D}$ \citep[][]{dentz2011JCH}. The angular brackets $\langle
\cdot \rangle$ denote the average over all noise realizations and
the particle position is represented by $\mathbf x(t|\mathbf a)$
with the initial position $\mathbf x(t = t_{0}|\mathbf a) = \mathbf
a$. Within the context of particle motion, $\rho(\mathbf a)$
corresponds to the distribution of initial particle positions. The
solute distribution $c(\mathbf x, t)$ is expressed in terms of the
particle trajectories $\mathbf x(t|\mathbf a)$ as
\begin{align}
c(\mathbf x,t) = \int \langle \delta[\vx - \vx(t|\mathbf a)] \rangle
\rho(\mathbf a) d \mathbf a.\label{eq:conc-rho}
\end{align}
In the following, we denote the Green function $\langle \delta[\vx -
\vx(t|\mathbf a)] \rangle$ by
\begin{align}
g(\vx,t|\va) = \langle \delta[\vx - \vx(t|\mathbf a)] \rangle.
\end{align}

The flow and transport equations~\eqref{eq:PDE-H}--\eqref{langevin}
fully describe passive solute transport in the heterogeneous porous
medium characterized by the spatially varying hydraulic conductivity
$K(\vx)$. Once $\mathbf v(\vx)$ is obtained from~\eqref{eq:PDE-H},
solute transport can be quantified by solving the
advection-dispersion equation~\eqref{eq:PDE-C} or the Langevin
equation~\eqref{langevin}. The numerical solution of
\eqref{eq:PDE-C} on a discretized flow domain consisting of grid
blocks of characteristic size $\lambda$ implies the loss of velocity
variability on scales below $\lambda$. The same applies to the
concentration distribution and particle trajectories that are
obtained from a smoothed, e.g. interpolated, flow field. Only in the
case of a discretization scale of the order of the smallest
heterogeneity scale, i.e. a highly refined numerical mesh, does the
numerical solution reproduce the full heterogeneity induced flow and
transport behavior. However, for a fine discretization of the flow
domain, the computational burden is heavy, specially in the context
of stochastic simulations where the Monte Carlo framework is
commonly adopted \citep[e.g.,][]{leube2013}. Thus, limited
computational resources impose a constraint on the design of the
numerical mesh and coarse grid blocks are generally used. On the
other hand, it is generally not possible to have detailed knowledge
on the fluctuation behavior of $K(\mathbf{x})$ over various orders
of magnitude. As a consequence, the resolution of $K(\mathbf{x})$
relies on a given practical characterization scale (for example, a
measurement grid) and the spatial resolution of the measurement
device \citep{beckie1996}.

Thus, it is necessary to account for the impact of velocity
fluctuations below the grid or characterization scale $\lambda$ on
large scale transport. Solute transport in the coarse scale
$\lambda$ renders concentration distributions and particle
trajectories that are smoother than the ones obtained from the fully
resolved fine scale velocity field. This leads to an inaccurate
description of dispersion and large scale transport
\cite{ababou1989,rubin1999,efendiev2000,deBarros2011jfm} with
implications in human health risk estimation and decision making
\citep{deBarros2009WRR,tartakovsky2013}. The challenge lies in
compensating the random subscale velocity variability, that is
homogenized at a scale smaller than $\lambda$, through its impact of
effective solute dispersion. We approach this problem using
stochastic modeling in conjunction with a spatial filtering
methodology in order to quantify subscale velocity fluctuations in a
systematic manner.

\subsection{Stochastic Model}\label{subsec:stochastic_model}

The spatially varying hydraulic conductivity field is considered a
realization of a stationary and ergodic random field. Specifically,
the log-hydraulic conductivity, $Y\left(\mathbf{x}\right) =
\ln\left[K\left(\mathbf{x}\right)\right]$, is modeled as a
multi-Gaussian random space function \cite[][]{rubin2003book}. The
randomness in the medium properties is mapped onto the flow velocity
$\mathbf v(\vx)$ through the Darcy equation~\eqref{eq:PDE-H}, which
then is passed on to solute and particle transport
via~\eqref{eq:PDE-C} and~\eqref{langevin}.

The multi-Gaussian $Y(\vx)$ is characterized by its constant mean
$\overline{Y(\vx)} = \overline{Y}$, variance $\sigma_{Y}^{2}$,
correlation length scale $l_i$ (with $i=1,...,d$) and two-point
spatial covariance
$C_{Y}\left(\mathbf{r}\right)=\overline{Y^{\prime}\left(\mathbf{x}\right)Y^{\prime}
  \left(\mathbf{x}+\mathbf{r}\right)}$ where $\mathbf{r}$ is the lag distance between two distinct
locations within the flow domain and $Y^{\prime}(\vx) = \overline{Y}
- Y^\prime(\vx)$ is the log-conductivity fluctuation. The overbar
denotes the ensemble average over all realizations of $Y(\vx)$. In
this work, the Fourier transform of $Y^\prime(\vx)$ is denoted by
$\widetilde Y^\prime(\vk)$ where $\vk$ represents the wave number.
The Fourier transform of a function $\varphi(\vx)$ is defined by
\begin{align}
\widetilde{\varphi} \left(\mathbf{k}\right) &=
\int_{-\infty}^{\infty}\exp\left(\imath
\mathbf{k}\cdot\mathbf{x}\right)
\varphi\left(\mathbf{x}\right)d\mathbf{x},
\\
\varphi\left(\mathbf{x}\right) &= \int_{-\infty}^{\infty}
\exp\left(-\imath \mathbf{k}\cdot\mathbf{x}\right)
\widetilde{\varphi}
\left(\mathbf{k}\right)\frac{d\mathbf{k}}{\left(2\pi\right)^{d}}.
\end{align} where $\imath$ is the imaginary unit.
Here and in the following, Fourier transformed quantities are marked
by a tilde. Furthermore, we employ the short-hand notation
\begin{align}
\int_k [\cdot] \equiv \int\limits_{-\infty}^\infty [\cdot] \frac{d
\vk}{(2
  \pi)^d}.
\end{align}

By making use of Darcy's law and representing the conductivity by a
Taylor series expansion about its mean value $\overline{Y}$, we can
obtain an approximation for the steady state velocity field
\cite{rubin1990wrr,rubin2003book}. First-order perturbation theory
in the log-conductivity fluctuations gives for the velocity field
$\mathbf v(\vx)$
\begin{align}
\label{vpt} v_i(\vx) = \overline v \delta_{i1} + \overline{v}
\int\limits_{-\infty}^\infty \exp(-i \vk \cdot \vx) p_i(\vk) \tilde
Y(\vk) \frac{d \vk}{(2\pi)^d},
\end{align}
where $\vk$ is the wave number. We assume that the mean hydraulic
gradient is aligned with the $1$--direction of the coordinate
system, $\overline{\nabla h} = G \mathbf e_1$. The mean velocity is
$\overline v = G K_g$, where $K_g = \exp(\overline{Y})$ corresponds
to the geometric mean hydraulic conductivity. The projectors
$p_i(\vk)$ are defined by $p_i(\vk) = \delta_{i1} - k_1 k_i/\vk^2$
and guarantee that $\nabla \cdot \mathbf v(\vx) = 0$ through
$\sum_{i=1}^d k_i p_i(\vk) = 0$.

The velocity field~\eqref{vpt} is a linear functional of the
multi-Gaussian $Y^\prime(\vx)$ and therefore, corresponds to a
multi-Gaussian random space function itself. Its mean is given by
$\overline{\mathbf v(\vx)} = \overline v \mathbf e_1$ and the
velocity fluctuations $v_i^\prime(\vx) = v_i(\vx) - \overline v
\delta_{i1}$ are expressed in the second term on the right hand side
of~\eqref{vpt}. The covariance of their Fourier transforms is
\begin{align}
\overline{\widetilde v^\prime_i(\vk) \widetilde
v^\prime(\vk^\prime)} = (2 \pi)^2 \delta(\vk + \vk^\prime)
\widetilde C_{ij}(\vk), && \mathrm{with} && \widetilde C_{ij}(\vk) =
p_i(\vk) p_j(\vk) \widetilde C_Y(\vk).\label{eq:vel-cov}
\end{align} where $\widetilde C_{ij}(\vk)$ is the velocity covariance
function in Fourier space. The Dirac delta in equation
(\ref{eq:vel-cov}) is a consequence of the stationarity of the
random field $Y(\vx)$.
%
\subsection{Solute Transport at the Blockscale}\label{subsec:blockscale_dispersion}
%
Solute transport is investigated from the Lagrangian point of view
starting from the Langevin equation~\eqref{langevin}. We begin by
defining a coarse grained velocity field and the corresponding
subscale velocity fluctuations, which in this framework are
incorporated into a suitably defined mesoscopic noise term.

\begin{figure}
\includegraphics[width=.9\textwidth]{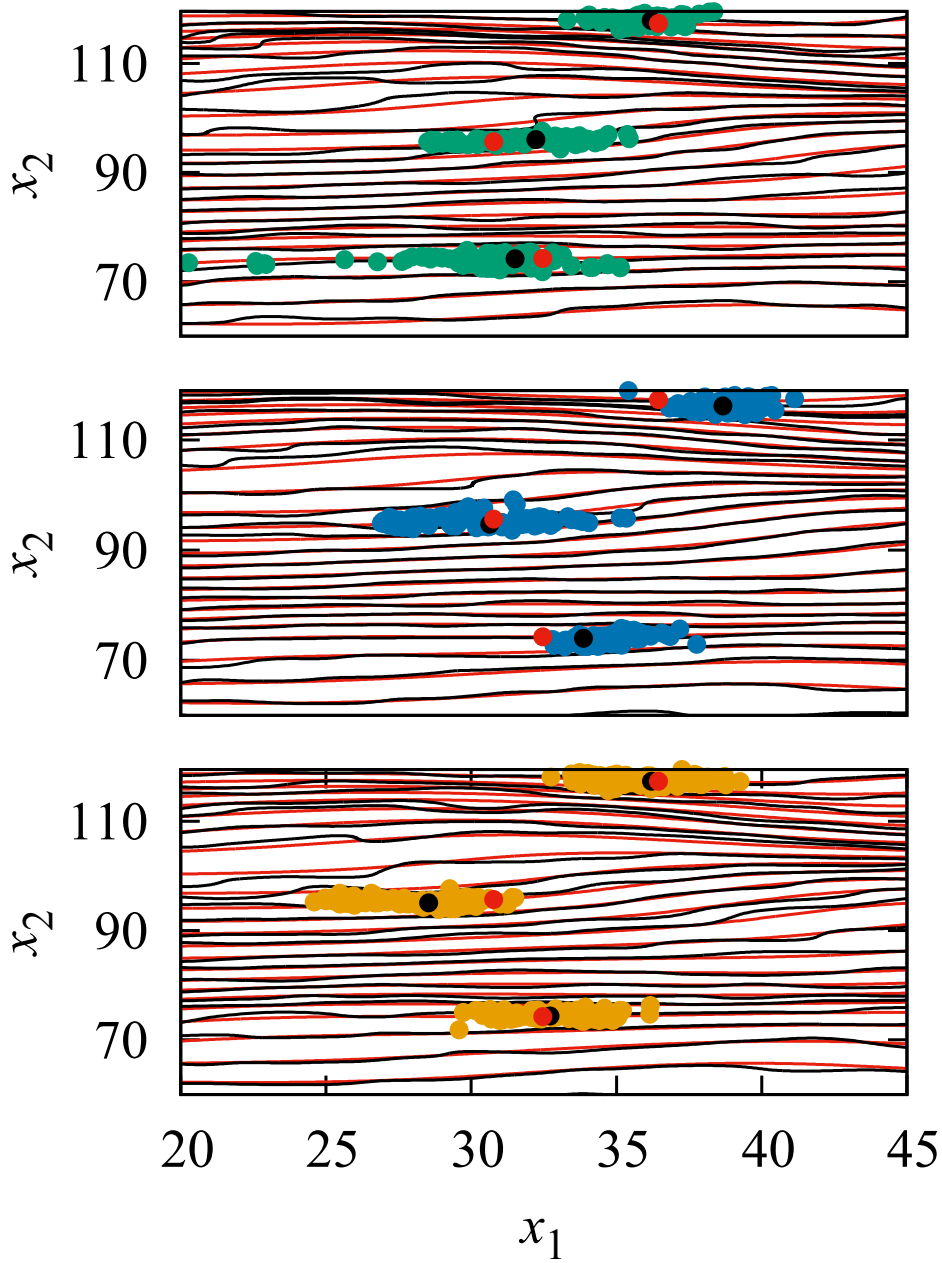}
\caption{Illustration of particles plumes in different realizations of
  $\mathbf v(\mathbf x)$ characterized  by the same coarse-grained
  $\mathbf v_>(\mathbf x)$. The solid black lines illustrate the
  streamlines of the fully variable velocity field $\mathbf v(\mathbf
  x)$ and the solid red lines the streamlines of the coarse-grained
  velocity field $\mathbf v_>(\mathbf x)$. The black dots denote the
  center of mass of the respective local particle distribution in the
  individual realization, while the red dots show the center of mass
  position obtained from the coarse-grained velocity $\mathbf
  v_>(\mathbf x)$.    \label{fig:plumes}}
\end{figure}

The spatial velocity fluctuations are separated in the contributions
below and above the coarse graining scale $\lambda$ in order to
determine the dispersion effect due to subscale velocity fluctuations.
The coarse-grained velocity field $\mathbf{v}_{>}(\mathbf{x})$ is
defined by spatial filtering as
\begin{align}
\label{vcoarse}
\mathbf{v}_{>}(\mathbf{x})=\int
d\mathbf{r} \mathcal F(\mathbf{x}-\mathbf{r})\mathbf{v}(\mathbf{r}),
\end{align}
with $\mathcal F(\mathbf{x})$ corresponding to a suitably chosen
spatial filter \cite{cushman1984,baveye1984,quintard1994,
beckie1996theory,rubin1999,attinger2003,dentz_debarrosJFM15}. This
coarse-graining operation reads in Fourier space as
\begin{align}
\widetilde{\mathbf{v}}_{>}({\mathbf k})=\widetilde{\mathcal
F}(\mathbf{k})\widetilde{\mathbf{v}}(\mathbf{k}),
\end{align}
where the spatial filter $\mathcal F(\vx)$ is normalized, i.e.,
\begin{align}
\int \mathcal F(\vx) d \vx = 1, && \widetilde{\mathcal F}(\vk =
\mathbf 0) = 1.
\end{align}
The subscale scale velocity fluctuations are defined correspondingly
by
\begin{align}
\mathbf{v}_{<}(\mathbf{x}) =
\mathbf{v}(\mathbf{x})-\mathbf{v}_{>}(\mathbf{x}).
\end{align}
Thus, the ensemble mean of the coarse grained flow
velocity~\eqref{vcoarse} is equal to the mean flow velocity while
the subscale fluctuations are, in average, zero,
\begin{align}
\overline{\mathbf{v}_{>}(\mathbf{x})}=\overline{\mathbf{v}}, &&
\overline{\mathbf{v}_{<}(\mathbf{x})}=\mathbf{0}\label{eq:avg-vel}
\end{align}
Figure~\ref{fig:plumes} illustrates three random velocity fields
$\mathbf v(\vx)$ characterized by the same coarse-grained velocity
field $\mathbf v_>(\mathbf x)$.

Next, we rewrite the fine scale Langevin equation~\eqref{langevin}
in the following form
\begin{align}
\frac{d\mathbf{x}(t|\va)}{dt} = \mathbf{v}_{>}[\mathbf{x}(t|\va)] +
\boldsymbol{\zeta}(t|\va), \label{Langevinc}
\end{align}
where the mesoscale noise $\boldsymbol{\zeta}(t|\va)$ is defined in
terms of the subscale velocity fluctuations and the local noise term
as
\begin{align}
\label{zeta}
\boldsymbol{\zeta}(t|\va) = \mathbf{v}_{<}[\mathbf{x}(t|\va)]
+\sqrt{2 \mathbf D} \cdot \boldsymbol \xi(t).
\end{align}
The mesoscale noise $\boldsymbol{\zeta}(t|\va)$ represents the
impact of subscale heterogeneity on coarse-scale solute dispersion.
Notice that $\boldsymbol \zeta(t|\va)$ is a multi-Gaussian noise due
to the Gaussianity of $\boldsymbol \xi(t)$ and $\mathbf v(\vx)$
(i.e. stochastic process for the velocity fluctuations at the
subscale). Figure~\ref{fig:plumes} illustrates particle
distributions in different realizations of $\mathbf v(\vx)$ for the
same coarse-grained $\mathbf v_>(\mathbf x)$. The effective noise
quantifies fluctuations about the coarse-grained velocity $\mathbf
v_>(\mathbf x)$. Note that the center of mass of the particle
distribution in individual realization does in general not coincide
with the coarse-grained center of mass (compare the locations of the
black dots and the red dots in Figure~\ref{fig:plumes}). In fact, as
discussed in the following, the statistical properties of
$\boldsymbol \zeta(t|\va)$ depend on the order in which the noise
and ensemble averages are performed. In the upcoming section, we
will show how the order of the averaging operation leads to distinct
conceptualizations of the blockscale dispersion tensor.

\section{Blockscale Dispersion Concepts\label{sec:dispersion_definition}}

In the previous section, we defined the mesoscale
noise~\eqref{zeta}, in order to quantify the dispersion effect due
subscale velocity fluctuations. It comprises local noise, which
models the impact of local dispersion, as well as the filtered
velocity fluctuations. Thus, $\boldsymbol \zeta(t|\va)$ depends on
the local noise, on the realization of the flow velocity as well as
on the initial position $\va$ of the respective solute particle. The
statistical properties of $\boldsymbol \zeta(t|\va)$, specifically,
its covariance, depend on the order in which the averages over the
noise and flow ensemble are taken, and which the integral over the
initial particle positions is performed. In the following, we
discuss two different averaging procedure, which define the
statistical properties of the mesoscale noise. These are as follows:

\begin{figure}
\includegraphics[width=\textwidth]{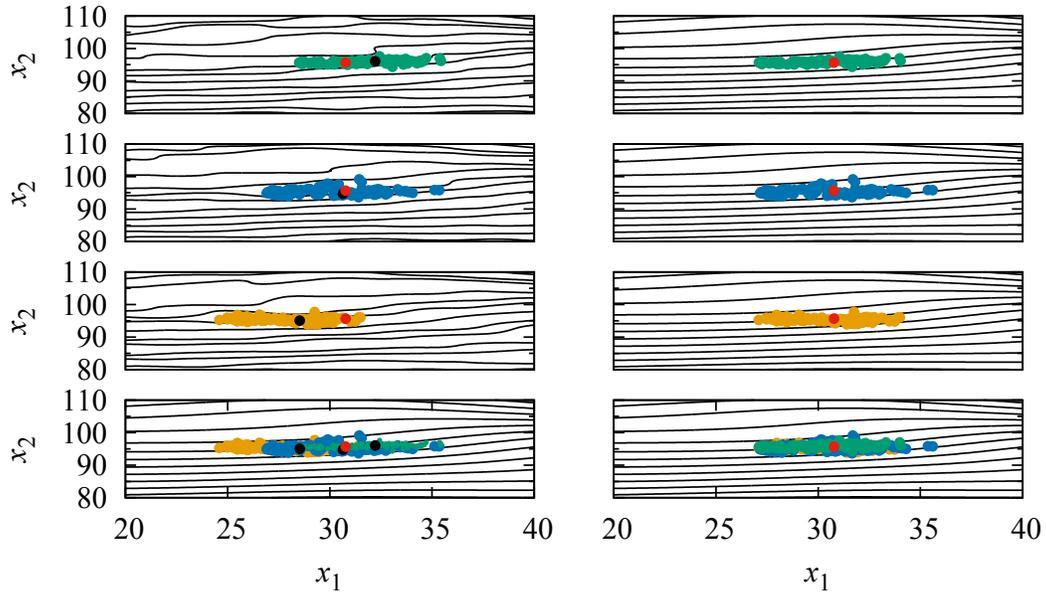}
\caption{Illustration of the concepts of ensemble and effective
  blockscale dispersion. Solid black lines illustrate streamlines. (Left panel) Evolution of solute plumes
 (rows 1-3 from top to bottom) in three different realizations of $\mathbf v(\vx)$ characterized by the same
  $\mathbf v_>(\vx)$. The red dot denote the center of mass position
  obtained from the coarse-grained velocity field and
  the black dots denote the actual center of mass positions in the
  particular realization of $\mathbf v(\mathbf x)$.  The bottom
  figure illustrates the ensemble blockscale dispersion concept, which
  quantifies fluctuations of the center of mass velocity in
  individual realizations with respect to the coarse-grained center of
  mass velocity. (Right panel) Solute plumes centered with respect to the coarse-grained center of
  mass position in the coarse-grained velocity field. The bottom figure illustrates the effective blockscale
  dispersion concept, which quantifies the velocity fluctuations with
  respect to the center of mass velocity in individual realizations.     \label{fig:concepts} }
\end{figure}

\begin{enumerate}

\item The first procedure is termed \textit{ensemble}
    noise average. It defines the covariance of $\boldsymbol \zeta(t|\va)$ with respect to its
noise and flow ensemble average. It quantifies velocity fluctuations
with respect to the coarse-grained velocity $\mathbf v(\mathbf x)$
between realizations of the finescale velocity fluctuations $\mathbf
v_<(\mathbf x)$. As illustrated in Figure~\ref{fig:concepts}, the
center of mass position of a particle distribution in a realization
of $\mathbf v(\mathbf x)$ does not necessarily coincide with
$\mathbf v_>(\mathbf x)$. Thus, the ensemble blockscale dispersion
quantifies an artificial ensemble effect due to center of mass
fluctuation between realizations of $\mathbf v_<(\mathbf x)$.
\item The second procedure is termed
    \textit{effective} noise average. It measures the dispersion
    effect of velocity fluctuations with respect to the actual mean
    velocity in a single realization of $\mathbf v_<(\mathbf x)$,
    which does not, in general, coincide with the coarse-grained
    $\mathbf v_>(\mathbf x)$, see also Figure~\ref{fig:concepts}. Thus the effective noise average
    represents genuine subscale velocity fluctuations.
\end{enumerate}

Based on these concepts, we define two surrogate processes that are
formalized by Langevin equations, which are characterized by the
respective ensemble and effective noises. These noises are
correlated Gaussian processes characterized by the respective
covariances. From these Langevin equations, we then derive the
corresponding advection-dispersion equations that govern the
evolution of the coarse-scale solute distributions in the two
different dispersion models.

\subsection{Blockscale Dispersion Based on the Ensemble Noise
Average}\label{subsec:ens}

The first averaging procedure employed and investigated is the
ensemble noise average of $\boldsymbol \zeta(t|\va)$. The ensemble
noise average of the mesoscale noise considers the covariance of the
mesoscale noise with respect to the ensemble noise average
\begin{align}
\overline{\langle \boldsymbol \zeta(t|\va) \rangle} =
\boldsymbol 0.
\end{align}
Thus, we define the
ensemble covariance of the mesoscale noise by
\begin{align}
C^{ens}_{ij}(t,t^{\prime}) = \overline{\langle \zeta_i(t|\va)
\zeta_j(t^\prime|\va) \rangle}.\label{eq:cov-ens}
\end{align}
The ensemble covariance quantifies the velocity covariance with
respect to the coarse-grained velocity $\mathbf v_>(\mathbf x)$. The
expression on the right hand side of~\eqref{eq:cov-ens} can be
expanded such that we obtain for the ensemble covariance
\begin{align}
{C}^{ens}_{ij}(t - t^{\prime}) &= 2 D_{ii} \delta_{ij}
\delta(t-t^{\prime}) + \int_{k}\int_{k^{\prime}}
\int_{k^{\prime\prime}} \mathcal A({\mathbf
k}) \mathcal A({\mathbf k}^\prime)
\nonumber\\
& \times
\overline{\widetilde{v}_{i}(\mathbf{k})\widetilde{v}_{j}(\mathbf{k}^{\prime})
 \widetilde{g}(-\mathbf{k},t-t^{\prime}|\va) \exp(i
 \vk^{\prime\prime} \cdot \va)
  \widetilde{g}(- \mathbf{k}^{\prime} - \mathbf{k}^{\prime\prime},t^{\prime}|\va)},
\label{cross}
\end{align}
where Lagrangian stationarity is assumed, $\mathcal A({\mathbf k}) =
1-\widetilde{\mathcal F}(\mathbf{k})$ and the function $\widetilde
g(\vk,t|\va)$ is the Fourier transform of the Green function
$g(\vx,t|\va) = \langle \delta[\vx - \vx(t|\va)] \rangle$. As we
will see in the following, the ensemble average covariance does not
depend on the initial particle position, as desired for an effective
framework.
%
%
The blockscale ensemble dispersion tensor can be evaluated by
integrating the ensemble covariance of the mesoscopic noise
$C_{ij}^{ens}$:
\begin{align}
D_{ij}^{ens}(t-t_0) = \int\limits_0^{t - t_0} d t^\prime
C_{ij}^{ens}(t^\prime), \label{Dens}
\end{align}
where $t_0$ is the initial time. The dispersion tensor
$D_{ij}^{ens}(t-t_0)$ in (\ref{Dens}) is also reported in eq. 4 of
Fiori et al. \cite{Fiori:2013}. Given the aforementioned
development, we may now define the coarse-grained stochastic process
\begin{align}
\frac{d \vx(t|\va)}{d t} = \mathbf v_>[\vx(t|\va)] + \boldsymbol
\zeta^{ens}(t).
\end{align}
where the ensemble noise $\boldsymbol \zeta^{ens}(t)$ is
characterized by zero mean $\langle \boldsymbol \zeta^{ens}(t)
\rangle_{ens} = \mathbf 0$ and the covariance function $\langle
\zeta^{ens}_{i}(t) \zeta^{ens}_j(t^\prime) \rangle_{ens}  =
{\mathcal C}^{ens}_{ij}(t - t^{\prime})$. The angular brackets with
subscript $ens$ denote the ensemble noise average. The average
concentration can be determined in terms of the ensemble noise
average as
\begin{align}
c^{ens}(\vx,t) = \int d \va \rho(\va) \langle \delta[\vx - \vx(t|\va)]
\rangle_{ens}.
\end{align}
Based on the fact that the mesoscale noise is Gaussian distributed,
one can derive the following advection-dispersion equation for
$c_{ens}(\vx,t)$
\begin{align}
\label{ADE:ens}
\frac{\partial c^{ens}(\vx,t)}{\partial t} + \mathbf v_>(\vx) \cdot
\nabla c^{ens}(\vx,t) - \nabla \cdot\left[ \D^{ens}(t - t_0) \nabla
  c^{ens}(\vx,t) \right] = 0.
\end{align}
Note that the dependence of $\D^{ens}(t - t_0)$ on the initial time
$t_0$ expresses the non-Markovian nature of the transport processes
due to the noise correlation~\cite{DLBC:2008, Fiori:2013}. For the
Langevin equation~\eqref{langevin}, which represents a Markov
process, the noise is delta-correlated, and thus the related
advection-dispersion equation~\eqref{eq:PDE-C} is independent of the
initial time.
\subsection{Blockscale Dispersion Based on the Effective Noise Average}\label{subsec:eff}
The effective noise average considers the covariance of the mesoscale
noise with respect to the velocity fluctuation averaged over the local
noise
\begin{align}
\langle \zeta_i(t|\va) \rangle = \int_k \mathcal A({\mathbf k})
\widetilde v_i(\mathbf{k}) \widetilde{c}(\mathbf{k},t|\va).
\end{align}
The effective single realization noise covariance is defined by
\begin{align}
\label{Ceffsingle}
{\mathcal C}^{eff}_{ij}(t,t^{\prime}) = \langle \zeta_i(t|\va)
\zeta_j(t|\va) \rangle - \langle \zeta_i(t|\va)
\rangle \langle \zeta_j(t^\prime|\va) \rangle.
\end{align}
This means it quantifies the velocity fluctuations with respect to
the mean velocity in single realization of the finescale velocity
$\mathbf v_<(\mathbf x)$, which is given by $\mathbf v_>(\mathbf x)
+ \langle \mathbf v_<(\mathbf x) \rangle$. Notice that the effective
noise covariance as an ensemble average does not depend on the
initial particle positions. Accordingly, we define now the effective
single realization dispersion coefficient as
\begin{align}
\label{Deffsingle}
\mathcal D_{ij}^{eff}(t) = \frac{1}{2} \int\limits_0^{t} d t^\prime
\left[\mathcal C_{ij}^{eff}(t,t^\prime) + \mathcal C_{ji}^{eff}(t,t^\prime)\right].
\end{align}
It measures the average relative solute and particle dispersion in a
single realization of $\mathbf v_<(\mathbf x)$. In this sense it may
be considered a \textit{measure for solute mixing} rather than the
ensemble dispersion coefficients, which quantify the uncertainty of
particle positions within the extended initial plume and between
heterogeneity realizations. Similar dispersion conceptualization can
be found in Fiori \cite{fiori:2001} and section 3.2 of Dentz and de
Barros \cite{dentz_deBarros2013}. The ensemble average
$C_{ij}^{eff}(t - t^\prime) = \overline{\mathcal
C^{eff}_{ij}(t,t^\prime)}$ can be written as
\begin{align}
{C}^{eff}_{ij}(t,t^{\prime}) &= {C}^{ens}_{ij}(t,t^{\prime}) - \int d
\va \rho(\va)
\nonumber\\
& \times
 \int_k \int_{k^\prime}
\mathcal A({\mathbf k}) \mathcal A({\mathbf k^\prime})
\overline{\widetilde v_i(\mathbf{k}) \widetilde
v_j(\mathbf{k}^\prime)
  \widetilde{g}(\mathbf{k},t|\va)
\widetilde{g}(\mathbf{k}^\prime,t^\prime|\va)}.
\end{align}
The blockscale effective dispersion coefficient now is given by
\begin{align}
D_{ij}^{eff}(t) = \int\limits_0^t d t^\prime
C_{ij}^{eff}(t^\prime) = \overline{\mathcal D^{eff}_{ij}(t)}.
\label{Deff}
\end{align}

In order to illustrate the difference between the ensemble and
effective dispersion coefficient, we consider the difference between
the noise average trajectories and its coarse-grained counter part,
$\delta \langle \vx(t|\va) \rangle = \langle \vx(t|\va) \rangle -
\langle \vx_>(t|\va) \rangle$, which is equal to
\begin{align}
\delta \langle \vx(t|\va) \rangle = \int\limits_0^t d t^\prime \langle
\boldsymbol \zeta(t^\prime|\va)\rangle.
\end{align}
It measures the deviation of the averaged coarse-grained particle trajectory
starting at $\va$ with the noise average of $\vx(t|\va)$. The
difference between ensemble and effective dispersion coefficients
measures the evolution of its variance
\begin{align}
\frac{[d \overline{\delta \langle x_i(t|\va) \rangle^2]_\rho}}{d t} =
D_{ii}^{ens}(t) - D_{ii}^{eff}(t).
\end{align}
The difference between the ensemble and effective dispersion
coefficients measures the evolution of the trajectory uncertainty
both due to fluctuations within the initial plume and between the
disorder realizations.

As above, we now define an effective coarse grained Langevin equation,
\begin{align}
\label{LangevinEff}
\frac{d \vx(t|\va)}{d t} = \mathbf v_>[\vx(t|\va)] + \boldsymbol
\zeta^{eff}(t),
\end{align}
where the effective noise is characterized zero mean, $\langle
\boldsymbol \zeta^{eff}(t) \rangle_{eff} = \mathbf 0$ and the
covariance $\langle \zeta_i^{eff}(t) \zeta_j^{eff}(t^\prime)
\rangle_{eff} = C_{ij}^{eff}(t - t^\prime)$. The angular brackets
with subscript \textit{eff} denote the effective noise average. We
define now an effective concentration in terms of the ensemble noise
average as
\begin{align}
c^{eff}(\vx,t) = \int d \va \rho(\va) \langle \delta[\vx - \vx(t|\va)]
\rangle_{eff},
\end{align}
where the angular brackets with subscript \textit{ens} denotes the
ensemble noise average. Based on the fact that the mesoscale noise
is Gaussian distributed one can derive the following
advection-dispersion equation for $c_{eff}(\vx,t)$
\begin{align}
\frac{\partial c^{eff}(\vx,t)}{\partial t} + \mathbf v_>(\vx) \cdot
\nabla c^{eff}(\vx,t) - \nabla \cdot\left[ \D^{eff}(t -
  t_0) \nabla c^{eff}(\vx,t) \right] = 0.
\end{align}
As before, the dependence of $\D^{eff}(t - t_0)$ on the initial time
is a manifestation of the non-Markovianity of~\eqref{LangevinEff} due
to the correlated noise.
\subsection{Self-Averaging of Dispersion}
The stochastic approach renders dispersion coefficients as averages
over an ensemble of realizations. Clearly, the dispersion behavior
in principle depends on the specific disorder characteristics. This
means, that average dispersion coefficients are in general not
representative of dispersion in single flow realizations. The
fluctuation of the solute plume spreading about its mean value is a
measure of heterogeneity-induced uncertainty
\cite{dagan1990,fiori2005awr,cirpka2011}. However, as more and more
of the flow heterogeneity is sampled by the solute plume, e.g., as
the transport histories experienced by the scalar in different
realizations, converge, the ensemble average is expected to become
representative of dispersion in a single realization
\citep{deBarros2011jfm,dentz_deBarros2013}. If the variance of the
dispersion coefficient tends to zero with time, then the dispersion
observable is self-averaging \cite{bouchaud1990}. More details on
self-averaging behavior of solute spreading in spatially
heterogeneous porous media can be found in the literature
\cite{dagan1990,deBarros2011jfm,dentz_deBarros2013,clincy2001,suciu2006,suciu2014,suciu2006firstorder,eberhard2007}.

The self-averaging behavior \cite[][]{bouchaud1990} is measured by
the variance of the single realization dispersion coefficients and
their ensemble averaged counterparts. Thus, we have for the
effective dispersion coefficients the variances
\begin{align}
\overline{\delta \mathcal D^{eff}_{ij}(t)^2} &= \overline{\left[\mathcal
    D^{eff}_{ij}(t) - D_{ij}^{eff}(t) \right]^2}.
\label{VarDeff}
\end{align}
Expressions for the fully upscaled (e.g. $\lambda \rightarrow
\infty$) dispersion variance can be found in Dentz and de Barros
\cite{dentz_deBarros2013} and asymptotic dispersion variance results
based on work of Fiori \cite{fiori1998} are also reported in
Appendix C of Wood et al. \cite{wood2003}. In the following section,
we develop explicit perturbation theory expressions for the ensemble
and effective blockscale dispersion coefficients as well as the
dispersion variances.
\section{Expressions for the Blockscale Dispersion Mean
  and Variance}
%
\subsection{Perturbation Theory}
The perturbation theory expressions for the ensemble and effective
dispersion coefficients are obtained straightforwardly
from~\eqref{Dens} and~\eqref{Deff} by substituting $\widetilde
g(\vk,t|\va)$ by
\begin{align}
\widetilde{g}_0(\vk,t|\va) = \exp(- \vk \cdot \D \vk + \imath
\overline{v} k_1 t) \exp(\imath \vk \cdot \va).
\end{align}
with $\imath$ corresponding to the imaginary unit. After some
algebraic manipulations, we obtain the expressions for the mean
dispersion coefficients:
\begin{align}
\label{DensPT} D_{ij}^{ens}(t) &= D_{ii} \delta_{ij} +
\int\limits_0^t dt^\prime \int_{k} \mathcal A({\mathbf k}) \mathcal
A(-{\mathbf k}) \widetilde C_{ij}(\vk)
  \widetilde{g}_0(-\mathbf{k},t-t^{\prime}|\mathbf 0),
\\
D_{ij}^{eff}(t) &= D_{ij}^{ens}(t) - \int\limits_0^t dt^\prime
\int_{k} \mathcal A({\mathbf k}) \mathcal A(-{\mathbf k}) \widetilde
C_{ij}(\vk) \widetilde{g}_0(\mathbf{k},t|\mathbf 0)
\widetilde{g}_0(-\mathbf{k},t^{\prime}|\mathbf 0). \label{DeffPT}
\end{align} with $\widetilde C_{ij}$ given in equation (\ref{eq:vel-cov}).
In order to determine perturbation theory expressions for the
dispersion variances, see~\eqref{VarDeff}, we need to expand the
single realization quantities up to first order in the velocity
fluctuations. To this end, we consider~\eqref{Deffsingle} as well as
the expression~\eqref{Ceffsingle} for the corresponding covariance
functions. Up to first order in the velocity fluctuations, we obtain
explicitly,
\begin{align}
\mathcal D_{ij}^{eff}(t) &= D_{ii} \delta_{ij}
\nonumber\\
& - \int\limits_k \imath  g_0(\vk,t|\mathbf 0) \widetilde \rho(\vk)
\mathcal A(\vk) \left[ \widetilde v_i(\vk) D_{jj}
  k_j t+ \widetilde v_j(\vk) D_{ii} k_i t
 \right].
\end{align} where $\imath$ is the imaginary unit.
Explicit result (in second order perturbation theory in the velocity
fluctuations) for the dispersion variance is provided below
\begin{align}
\overline{\delta \mathcal D_{ij}(t)^2} &= \int\limits_k
g_0(\vk,t|\mathbf 0) g_0(-\vk,t|\mathbf 0) \widetilde \rho(\vk)
\mathcal A(\vk) \widetilde \rho(-\vk) \mathcal A(-\vk)
\nonumber\\
& \times \left[\widetilde{C}_{ii}(\mathbf{k})
  D_{jj}^2k_j^2 t^2 + \widetilde{C}_{jj}(\mathbf{k}) D_{ii}^2 k_i^2 t^2 + 2 \widetilde{C}_{ij}(\mathbf{k})
  D_{ii} k_i D_{jj} k_j t^2 \right].\label{eq:gen-variance}
\end{align} where $A(\vk)=1-\widetilde{\mathcal{F}}(\vk)$ and
$\widetilde{\mathcal{F}}(\vk)$ is the Fourier transform of the
filter function (see section \ref{subsec:stochastic_model}) and
$\widetilde C_{ij}$ is defined in equation (\ref{eq:vel-cov}).
Details related to the derivation of the unfiltered dispersion
variance can be found in Dentz and de Barros
\cite{dentz_deBarros2013}. As opposed to the analysis of
\cite{dentz_deBarros2013}, equation~\eqref{eq:gen-variance}
incorporates the effect of the blockscale $\lambda$ on the
sample-to-sample fluctuations of the dispersion behavior.
Furthermore, contrary to the blockscale dispersion variance derived
in de Barros and Rubin \cite{deBarros2011jfm},
equation~\eqref{eq:gen-variance} accounts for the effects of local
scale dispersion.
%
\subsection{Analytical and Semi-Analytical Solutions}
In the following, analytical and semi-analytical expressions for the
average and variance dispersion coefficients are developed. These
expressions allows to obtain important physical insight of the
temporal evolution of the dispersion coefficients and its
self-averaging behavior. Without loss of generality, we employ the
square filter (in Fourier space)
\begin{eqnarray}
\widetilde{\mathcal{F}}\left(\vk \right) & = &
\begin{cases}
1 & \mathrm{for}\:\left|k_{i}\right|\leq\frac{\pi}{\lambda_{i}}\\
0 & \mathrm{otherwise}
\end{cases}\label{eq:filter}
\end{eqnarray}
where $i=$1,...,$d$, and $\lambda_{i}$ is the block size in the i-th
direction. The filter $\widetilde{\mathcal{F}}\left(\vk \right)$ has
been used to investigate the mean blockscale dispersion behavior
\cite{rubin1999,rubin2003wrr,bellin2004SERRA,eberhard2005upscaling,lawrence2007}
and the variance and probability density function of the blockscale
dispersion tensor within a purely advective stratified random flow
field \cite{deBarros2011jfm}. The $Y=\ln K$ covariance model
selected for our analysis is the Gaussian model

\begin{eqnarray}
C_{Y}(\mathbf{x}-\mathbf{x^{\prime}}) & = & \sigma_{Y}^{2}
\prod_{i=1}^{d} \exp\left[-\frac{\mathbf{\left\|\mathbf{x}
        - \mathbf{x^{\prime}}\right\|}^{2}}{2l_{i}^{2}}\right],\label{eq:Gauss-cov-anisot}
\end{eqnarray} where $\|\ cdot \|$ is the vector norm. Other correlation models for $Y$ can also be
employed.
Furthermore, without loss of generality, we employ a Gaussian
distribution for the initial condition $\rho\left(\mathbf{x}\right)$
with characteristic dimensions $L_i$ (with $i=1,...,d$):
\begin{eqnarray}
\rho\left(\mathbf{x}\right) & = & \prod_{i=1}^{d}\frac{1}{\sqrt{2\pi
L_{i}^{2}}}\exp\left[-\frac{x_{i}^{2}}{2L_{i}^{2}}\right].\label{eq:source-dist}
\end{eqnarray} Equation \eqref{eq:source-dist} allows to obtain analytical
solution for the blockscale dispersion coefficients and reduces to a
point-like injection in the limit $L_i$ $\rightarrow$ $0$. Other
types of source zone distributions can be adopted and the
computation of the blockscale dispersion coefficients might require
numerical integrations.
The quantities above, namely \eqref{eq:Gauss-cov-anisot} and
\eqref{eq:source-dist}, are expressed in Fourier space as

\begin{eqnarray}
\widetilde{C}_{Y}\left(\mathbf{k}\right) & = & \sigma_{Y}^{2}
\left(2\pi\right)^{d/2} \prod_{i=1}^{d} l_{i}
\exp\left[-\frac{1}{2}\left(k_{i}^{2}l_{i}^{2}\right)
\right],\label{eq:cov-K}\\
\widetilde{\rho}\left(\mathbf{k}\right) & = &
\prod_{i=1}^{3}\exp\left[-\frac{k_{i}^{2}L_{i}^{2}}{2}\right].\label{eq:rho-FT}
\end{eqnarray}

%
\subsubsection{Average Dispersion Coefficients}

Closed-form expressions for the ensemble and effective dispersion
coefficients are presented in the limit of large P\'{e}clet, $Pe \gg
1$, and $t \gg \tau_v$. The P\'{e}clet number is defined as $Pe
\equiv \tau_{D_{1}}/ \tau_v$ where $\tau_v$ is the advection time
scale defined by $\tau_v = l_1 / \overline v$ (i.e. characteristic
advection time over a longitudinal correlation length $l_1$) and
$\tau_{D_i}$ the dispersion time scale given by $\tau_{D_i} = l_i^2
/ D_i$.

The off-diagonal elements of the average dispersion coefficients are
zero for symmetry reasons. Furthermore, in first-order perturbation
theory in the fluctuation variance of $Y$, only the longitudinal
dispersion coefficients are of macroscopic order, i.e., their
leading order is independent of the local scale dispersion
coefficients. Thus, we focus here only on the longitudinal
dispersion coefficients. Substituting~\eqref{eq:filter},
~\eqref{eq:cov-K} and~\eqref{eq:rho-FT} in~\eqref{DensPT}, we obtain
the ensemble dispersion coefficient
\begin{align}
D_{11}^{{ ens}}(t\rightarrow\infty) \equiv D_{11}^{{ ens,\infty}} &=
D_{11} + \sqrt{\frac{\pi}{2}} \sigma_Y^2 \overline v l_1 \left[1 -
\prod_{i=2}^d {\rm erf}\left(\frac{\pi}{\sqrt{2} \lambda_i} \right)
\right], \label{eq:Densexpl}
\end{align}
which is similar to the asymptotic expressions derived in Rubin et
al. \cite{rubin1999,rubin2003wrr}. For the effective dispersion
coefficient we obtain in analogy
\begin{align}
D_{11}^{{ eff}}(t) &=  D_{11}^{{ ens,\infty}} - \sqrt{\frac{\pi}{2}}
\sigma_Y^2 \overline v l_1 \frac{1 - \prod_{i=2}^d {\rm
    erf}\left[\frac{\pi \left(1 + \frac{4
        t}{\tau_{D_i}}\right) }{\sqrt{2} \lambda_i } \right]}{\prod_{i = 2}^d \left(1
    + \frac{4 t}{\tau_{D_i}}\right)^{1/2}}.
\label{eq:DensEFFl}
\end{align}
%

\subsubsection{Dispersion Variance for a Fully Isotropic Case}

The dispersion variance, equation (\ref{eq:gen-variance}), can
evaluated through numerical quadratures. However, a semi-analytical
expression can be obtained by considering a fully isotropic case: an
isotropic source ($L=L_{i}$), a statistically isotropic $Y$-field
($l\equiv l_{i}$), an isotropic local scale dispersion ($D_{i}=D$)
and a cubic block of dimensions $\lambda$. The semi-analytical
solution is obtained by making use of the filter (\ref{eq:filter}),
the correlation function (\ref{eq:Gauss-cov-anisot}) and the inlet
distribution (\ref{eq:source-dist}). We report the results for a
three-dimensional flow field ($d=3$). The dispersion variance that
complies with the aforementioned simplifying assumptions is

\begin{align}
\overline{\delta \mathcal D_{ii}(t)^2} &=
\frac{8}{35}\sigma_{Y}^{2}l^{2}\overline{v}^{2}\frac{\left[(L/l)^{2}+\frac{2t}{\tau_{D}}\right]^{2}}{\left[1+2(L/l)^{2}+\frac{4t}{\tau_{D}}\right]^{5/2}}
- \frac{4 D^2 t^2 \gamma}{(2 \pi)^3} \int\limits_{\mu=0}^{\infty}
 \Omega\left(t,\mu\right)  d\mu .\label{eq:simp-variance}
\end{align} where $\gamma = \overline{v} \sigma_{Y}^{2} l^3 (2
\pi)^{3/2}$ and $\Omega\left(t,\mu\right)$ is:

\begin{eqnarray}
\Omega\left(t,\mu\right) & = & -\frac{4\sqrt{2} \pi\mu}
{\left[\lambda\Phi\left(t,\mu\right)\right]^{5}}
\exp\left[-\frac{3}{2}\frac{\pi^{2}}{\lambda^{2}}\Phi\left(t,\mu\right)\right]\Psi_{1}\left(t,\mu\right)\times\nonumber\\
&&\left[\Psi_{2}\left(t,\mu\right)+\Psi_{3}\left(t,\mu\right)-\Psi_{4}\left(t,\mu\right)\right]\label{eq:Omega-fxn}
\end{eqnarray}
with $\Phi$, $\Psi_{1}$, $\Psi_{2}$, $\Psi_{3}$ and $\Psi_{4}$
defined as

\begin{eqnarray}
\Phi\left(t,\mu\right) & = &
2\left(\frac{2l^{2}t}{\text{\ensuremath{\tau_{D}}}}+\mu+L^{2}\right)+l^{2}\label{eq:Phi-fxn}
\end{eqnarray}

\begin{eqnarray}
\Psi_{1}\left(t,\mu\right) & = & \lambda\sqrt{2\pi\Phi\left(t,\mu\right)}\text{erf}\left[\frac{\pi\sqrt{\Phi\left(t,\mu\right)}}{\lambda\sqrt{2}}\right]\exp\left[\frac{\pi^{2}\Phi\left(t,\mu\right)}{2\lambda^{2}}\right]-2\pi\Phi\left(t,\mu\right)\nonumber \\
\Psi_{2}\left(t,\mu\right) & = & \lambda^{3}\sqrt{\Phi\left(t,\mu\right)\pi}\exp\left[\frac{\pi^{2}\Phi\left(t,\mu\right)}{2\lambda^{2}}\right]\text{erf}\left[\frac{\pi}{\lambda}\sqrt{\frac{\Phi\left(t,\mu\right)}{2}}\right]\nonumber \\
\Psi_{3}\left(t,\mu\right) & = & \lambda\sqrt{\Phi\left(t,\mu\right)\pi}\left(2\pi^{2}\Phi\left(t,\mu\right)+4\lambda^{2}\right)\exp\left[\frac{\pi^{2}\Phi\left(t,\mu\right)}{2\lambda^{2}}\right]\text{erf}\left[\frac{\pi}{\lambda}\sqrt{\frac{\Phi\left(t,\mu\right)}{2}}\right]\nonumber \\
\Psi_{4}\left(t,\mu\right) & = &
2\sqrt{2}\lambda^{4}\exp\left[\frac{\pi^{2}\Phi\left(t,\mu\right)}{\lambda^{2}}\right]\text{erf}\left[\frac{\pi}{\lambda}\sqrt{\frac{\Phi\left(t,\mu\right)}{2}}\right]^{2}+\pi\sqrt{2}\lambda^{2}\Phi\left(t,\mu\right)\label{eq:Psi-i}
\end{eqnarray}

Summarizing, equation (\ref{eq:simp-variance}) contains one integral
that needs to be evaluated numerically. Notice that the following
limiting cases hold for equation (\ref{eq:simp-variance}):

\begin{eqnarray}
\lim_{\lambda\rightarrow\infty} \overline{\delta \mathcal D_{ii}(t)^2}
& = & \frac{8}{35}\sigma_{Y}^{2} l^{2}\overline{v}^{2}\frac{\left[(L/l)^{2}+\frac{2t}{\tau_{D}}\right]^{2}}{\left[1+2(L/l)^{2}+\frac{4t}{\tau_{D}}\right]^{5/2}};
\label{eq:limit}\\
\lim_{\lambda\rightarrow0}\overline{\delta \mathcal D_{ii}(t)^2} & =
& 0.\label{eq:limit2}
\end{eqnarray}
When the block scale is large compared to the heterogeneity
correlation scale, the blockscale dispersion variance tends to the
fully upscaled dispersion variance reported in Dentz and de Barros
\cite{dentz_deBarros2013} whereas for a very small block scale (e.g.
fine numerical grid), all the heterogeneity is captured in the
velocity field and the dispersion variance tends zero since the
dispersive fluxes are small. These limiting cases will be shown in
the upcoming section.

\section{Results and Analysis}

In this section we analyze the joint impact of the blockscale,
source injection scale, local scale dispersion and the statistical
anisotropy ratio in the overall mean effective dispersion behavior
and its fluctuations. In all simulations, we set $\overline{v}=1$,
$\sigma_{Y}^{2}=1$, $l_{i}=1$ and $\lambda=\lambda_{i}$ (for
$i=1,2,3$) unless stated otherwise. Note that $l_i$ corresponds to
the correlation scale of $Y$ and not the integral scale as adopted
in several other works \cite[e.g.][]{dagan1984,rubin1999}. For a
correspondence between the integral scale and the correlation scale
for the Gaussian covariance model \eqref{eq:cov-K}, see chapter 2 of
Rubin \cite{rubin2003book}. For the purpose of illustration, the
local scale dispersion tensor is set to be isotropic and components
equal to $D$.

\begin{figure}
\begin{center}
\includegraphics[width=.75\textwidth]{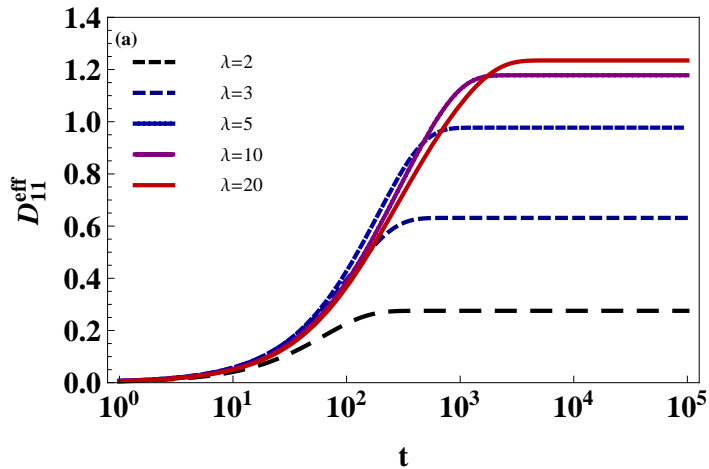}
\end{center}%
\caption{Temporal evolution of the average effective longitudinal blockscale
dispersion coefficient as a function of $\lambda$.
Analysis performed for a fully isotropic case and $Pe=10^{3}$.
\label{fig:mean-disp}}
\end{figure}

Figure \ref{fig:mean-disp} displays the average effective dispersion
coefficient (\ref{eq:DensEFFl}). Larger block sizes lead to larger
effective dispersion values. These results are expected and are in
agreement with previous analysis
\cite{rubin1999,rubin2003wrr,deBarros2011jfm}. Larger values of
$\lambda$ implies that a larger portion of the spatially variable
flow field is homogenized. Figure \ref{fig:mean-disp} also
illustrates the temporal evolution of the effective blockscale
dispersion coefficient $D_{11}^{eff}(t)$. It evolves roughly on the
dispersion time scale $\tau_D$. For decreasing ratio between block
and correlation scales, it evolves faster because the characteristic
heterogeneity scales that contribute to solute dispersion are below
the correlation scale and thus are sampled faster by the solute
plume (see also discussion in \cite{rubin2003wrr}). Notice that the
ensemble blockscale dispersion coefficient (\ref{eq:Densexpl})
evolves towards its asymptotic value on the advection scale
$\tau_v$. Thus, as for its unfiltered
counterpart~\cite[][]{kitanidis1988,dagan1990,dentz2000point,dentz2000ext},
at preasymptotic times, the ensemble blockscale dispersion
coefficients quantify artificial sample to sample fluctuations of
the center of mass position of the coarse-grained plume. They
overestimate actual solute dispersion due to subscale velocity
fluctuation. The effective blockscale dispersion coefficients on the
other hand reflect the dispersion effect due to local scale
dispersion and subscale velocity fluctuations only.

\begin{figure}[h]
\begin{centering}
\includegraphics[scale=0.4]{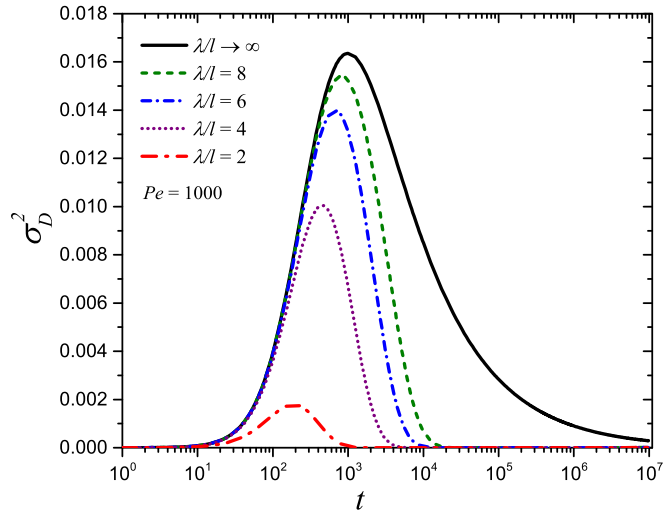}
\par\end{centering}
\centering{}\caption{Temporal evolution of the dispersion variance
as a function of the block scale $\lambda$. Results obtained for a
fully isotropic case, point source injection and $Pe=10^{3}$.}
\label{fig:variance-iso}
\end{figure}

Figure \ref{fig:variance-iso} depicts the temporal evolution of the
blockscale dispersion variance for a point source injection in a
fully isotropic scenario with $Pe=10^{3}$ where
$Pe\equiv\overline{v}l/D$. Notice that the dispersion variance is
given by $\sigma_{Dii}^2 \equiv$ $\overline{\delta \mathcal
D_{ii}(t)^2}$ in Figure \ref{fig:variance-iso}. The results were
obtained using the semi-analytical expression given by equation
(\ref{eq:simp-variance}). As expected, the dispersion variance
decreases with decreasing block size $\lambda$. For larger $\lambda$
values, more subscale variability is homogenized since less
heterogeneity is captured on the blockscale (such as in a coarse
numerical grid). Therefore, this wiped out variability needs to be
compensated in the dispersion tensor (see figure
\ref{fig:mean-disp}). For large $\lambda$, the longitudinal
blockscale dispersion variance (\ref{eq:simp-variance}) approaches
the fully upscaled value derived in \cite{dentz_deBarros2013}, see
the limit in equation (\ref{eq:limit}). Figure
\ref{fig:variance-iso} also shows that the values obtained for the
blockscale dispersion variance are small hence indicating that the
uncertainty is not large. From a practical point of view, this
result illustrates the potential of the blockscale, which is in
general defined by field sampling campaigns \cite{rubin1999}, in
reducing the uncertainty in transport predictions.

The other remarkable observation in figure \ref{fig:variance-iso} is
that the sample-to-sample fluctuations of the dispersion tends to
zero for large times. Therefore, dispersion is self-averaging.
However, by comparing the curve for $\lambda\rightarrow\infty$ with
the result obtained for $\lambda=2$, we observe that the
self-averaging behavior is anticipated for smaller block sizes. For
the fully upscaled dispersion tensor, i.e.
$\lambda\rightarrow\infty$, the system self-averages approximately
at time $t\approx 10^{7}$ while for $\lambda=2$, the system
self-averages around the diffusive time scale, i.e. $t\approx10^{3}$
This result has implications in field applications since, in
general, numerical grid design are dictated by a coarse hydraulic
conductivity measurement network. In other words, our results
indicate the significance of blockscale in reducing the variance of
the large scale plume spreading behavior. As shown in figure
\ref{fig:variance-iso}, conditioning model predictions on
conductivity measurements and other types of information has the
strong potential of reduce both the uncertainty of the large scale
dispersion behavior and the self-averaging characteristic time.

\begin{figure}[h]
\begin{centering}
\includegraphics[scale=0.4]{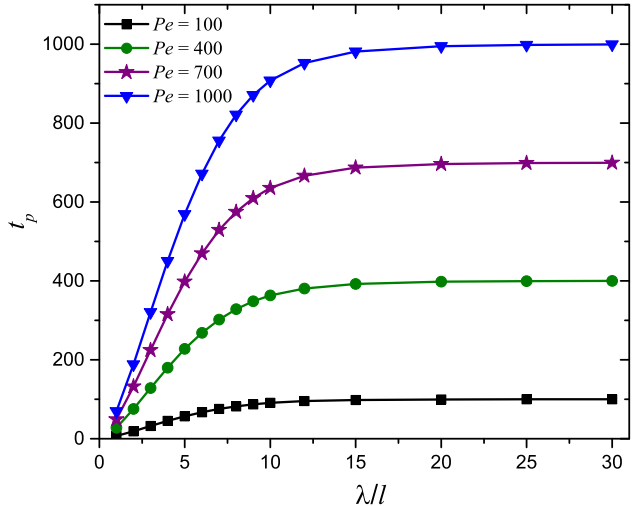}
\par\end{centering}
\begin{centering}
\caption{Time of the peak dispersion variance ($t_{p}$) versus the
dimensionless block scale $\lambda/l$. Results obtained for a fully
isotropic case, point source injection, $\overline{u}=1$,
$\sigma_{Y}^{2}=1$, $l=1$ and for different
$Pe$.}\label{fig:peak-time-iso}
\par\end{centering}
\end{figure}

Another key parameter in reducing the dispersion variance and
controlling the self-averaging rate of the spreading behavior of the
plume is the local scale dispersion coefficient $D$. Local-scale
dispersion tends to smooth out concentration gradients thus
diminishing the differences observed from the ensemble. Figure
\ref{fig:peak-time-iso} illustrates the dependency of the peak
dispersion variance time on the block size for different $Pe$. The
time of the peak dispersion variance is denoted by $t_{p}$ and
mathematically defined as

\begin{eqnarray}
t_{p}\left(\lambda\right) & = &
\arg\max_{t\in\left[0,\infty\right)}\sigma_{D11}^{2}\left(t;\,\lambda\right).
\end{eqnarray}

As shown in figure \ref{fig:peak-time-iso}, the time of peak
dispersion variance, $t_{p}$, will depend on $\lambda$. The time of
peak $t_{p}$ is more sensitive to $\lambda$ for higher $Pe$, e.g.
advection dominated scenarios. For lower $Pe$, $t_{p}$ reaches its
asymptotic value for smaller block sizes. Close inspection of figure
\ref{fig:peak-time-iso} shows that the asymptotic value for $t_{p}$
is equal to the diffusive characteristic time
$\tau_{D}=l^{2}/D_{d}$. For $Pe=10^3$, the asymptotic value for
$t_{p}$ is reached around $\lambda\approx 25l$ whereas for
$Pe=10^2$, $t_{p}$ approaches its asymptote at $\lambda\approx 10l$.

\begin{figure}
\begin{minipage}[t]{1\columnwidth}%
\begin{center}
\includegraphics[scale=0.35]{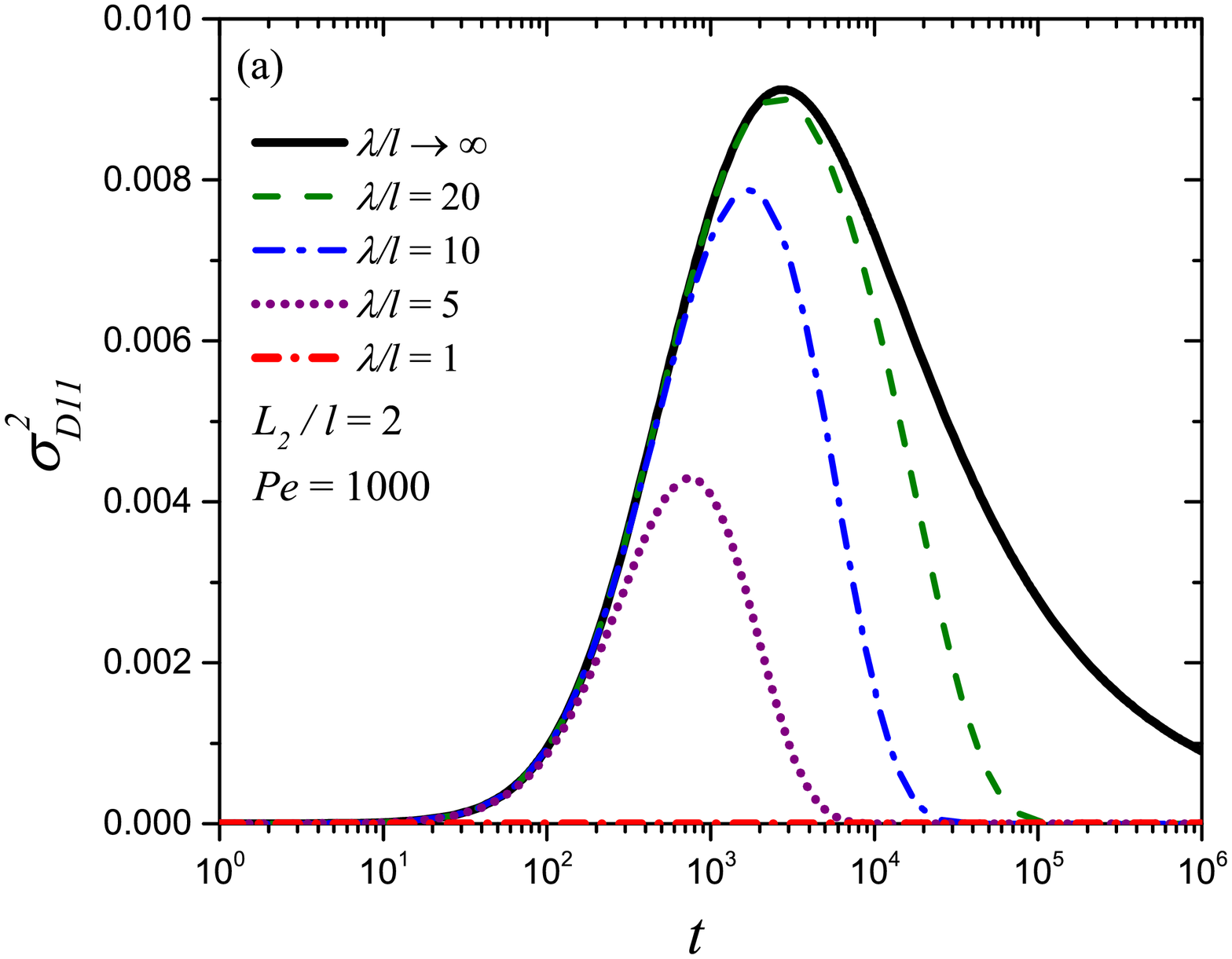}
\par\end{center}%
\end{minipage}
\vfill{}
\begin{minipage}[t]{1\columnwidth}%
\begin{center}
\includegraphics[scale=0.35]{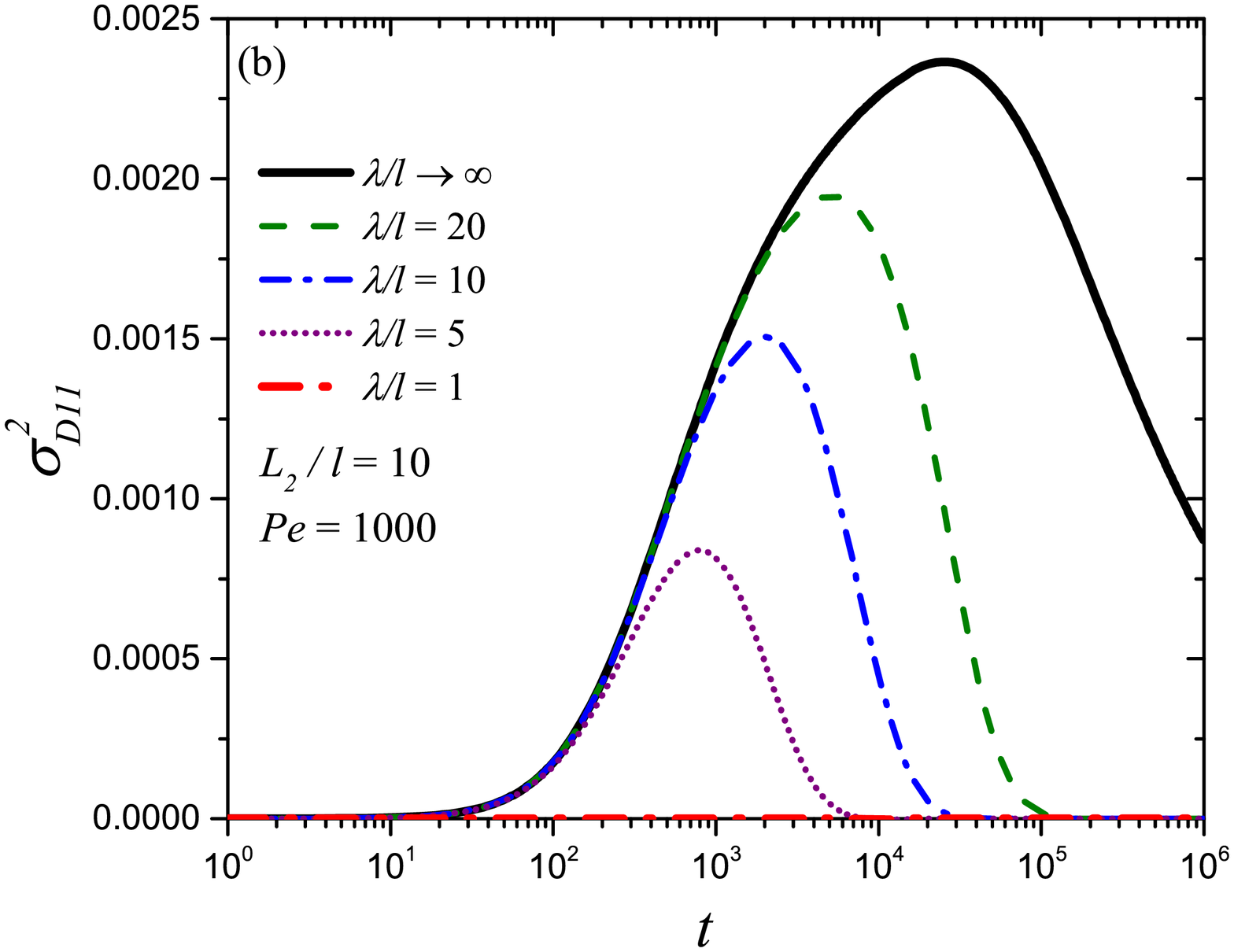}
\par\end{center}%
\end{minipage}
\caption{Temporal evolution of the dispersion variance as a function
of the block scale $\lambda$ for different transverse source
extension $L_{2}/l=\left[2, 10\right]$ with fixed $L_{1}=L_{3}=0.1$.
Results obtained for a fully isotropic case and
$Pe=10^{3}$.\label{fig:var-extended-line}}
\end{figure}

Next, we investigate the impact of the source dimension $L$ in
reducing the blockscale dispersion variance. The injection zone is
characterized by an extended transverse line source. Figure
\ref{fig:var-extended-line} shows the time evolution of the
dispersion variance for the following characteristic transverse
length scale of the source distribution: $L_{2}=2l$ and $L_{2}=10l$
for a fixed $L_{1}=L_{3}=0.1l$. As depicted in figure
\ref{fig:var-extended-line}, the sample-to-sample fluctuations of
the blockscale dispersion coefficient reduces with increasing
$L_{2}$. This effect can be observed by comparing figures
\ref{fig:var-extended-line}.a with \ref{fig:var-extended-line}.b.
For large injection sources, the solute plume approaches ergodicity
and transport is less subject to uncertainty
\cite{dagan1991,fiori1998,deBarros2011jfm}. The impact of $L$ on the
uncertainty of spreading was also investigated in detail in the
literature
\cite{zhang1996nonergodic,fiori2005awr,dentz_deBarros2013}.

\begin{figure}
\begin{centering}
\includegraphics[scale=0.45]{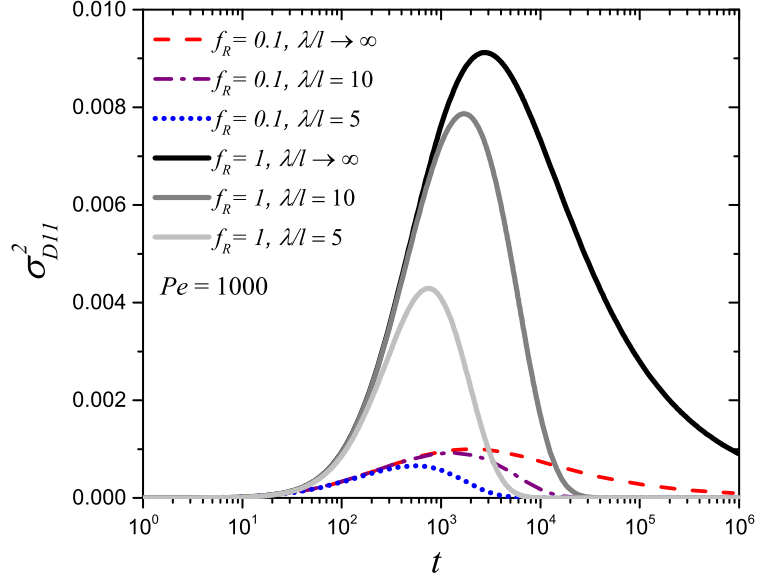}
\par\end{centering}
\centering{}\caption{Temporal evolution of the longitudinal
dispersion variance as a function of the block scale $\lambda$ and
the statistical anisotropy ratio $f_{R}$. Results for
$L_{1}=L_{3}=0.1l_{h}$, $L_{2}=2l_{h}$, $Pe=10^{3}$,
$f_{R}=l_{3}/l_{h}$ where $l_{h}\equiv l_{1}=l_{2}$. Parameter
values used $\overline{u}=1$, $\sigma_{Y}^{2}=1$ and
$l_{h}=1$.\label{fig:var-anisotropy-ratio}}
\end{figure}

The uncertainty of the large scale dispersion also depends on the
statistical anisotropy ratio $f_{R}$ of the log-conductivity. The
anisotropy ratio is defined as $f_{R}=l_{3}/l_{h}$, with
$l_{1}=l_{2}\equiv l_{h}$ denoting the horizontal correlation scale
and $l_{3}$ is the vertical correlation scale of $Y$. For many
geological sites, values of $f_R$ are typically $f_{R}\ll1$
\cite{rubin2003book}. Figure \ref{fig:var-anisotropy-ratio} displays
the impact of $f_{R}$ in reducing the blockscale dispersion
variance. The dispersion variance results are obtained for
$f_{R}=$0.1 and 1. For a smaller anisotropy ratio, see curves for
$f_{R}=0.1$, dilution is augmented since a lower value of $f_{R}$
tends to generate solute plumes with fingers of vertical
characteristic scale $l_{3}$. These thin solute fingers are smoothed
out by $D$ thus destroying the variability of the concentration
distribution in a single realization
\cite{kapoor1998,deBarros2014WRR}.

\section{Summary and Conclusions}

Numerical models tend to use coarse numerical grid blocks for two
main reasons. The first one is that the hydraulic conductivity and
flow data are, in general, obtained over a coarse measurement grid
and therefore the flow field is partially resolved. Second,
stochastic modeling of transport in complex hydro-systems often
calls for brute force numerical Monte Carlo simulations with a fine
numerical mesh such that solute spreading and mixing are accurately
predicted. This leads to a computational burden which can be
alleviated by using coarse numerical grid blocks. For such reasons,
it is important to wisely allocate computational resources
\cite{leube2013,moslehi2015}. A possible strategy consists in
accounting for the effects of flow variability at the subgrid scale
in dispersion coefficients. Two fundamental issues are pursued in
this article. We discuss two distinct blockscale dispersion
concepts, which represent different aspects of subscale velocity
fluctuation, and quantify the uncertainty in terms of the variance
of the sample to sample variance of the single realization
blockscale dispersion coefficients.

We discuss the concepts of ensemble and effective dispersion for the
quantification of subscale velocity fluctuations. As for its
unfiltered counterpart, effective dispersion reflects the dispersion
effect due to velocity fluctuations in a single aquifer realization.
The ensemble dispersion coefficient quantifies an artificial
dispersion effect due to sample to sample velocity fluctuations
between aquifer realizations. Thus, the effective blockscale
dispersion concept is better suited for the quantification of
dilution and chemical reactions within a coarse-grained transport
framework than the ensemble dispersion concepts. Based on these
blockscale dispersion conceptualizations, we propose two stochastic
surrogate models from which we derive the governing equations for
the respective coarse-grained concentration distributions.

The uncertainty of the blockscale dispersion tensor is only
well-defined for the effective blockscale dispersion concepts
because they are derived by the ensemble average over their single
aquifer realization counterpart. The ensemble dispersion coefficient
is genuinely an ensemble concept and defined in terms of the
ensemble of aquifer realizations as a whole. Uncertainty of the
blockscale dispersion coefficients is manifested through their
variance, which strongly depends on the initial dimensions of the
solute plume and the scale of the homogenization block.

We develop explicit (semi-) analytical solutions for the effective
blockscale dispersion coefficients and the blockscale dispersion
variance. Notice that our analytical and semi-analytical results are
limited to uniform-in-the-mean flow fields, unbounded domains,
low-to-mild heterogeneity ($\sigma^{2}_{Y} < 1$), a Gaussian
covariance function and an instantaneous injection of a non-reactive
solute following a spatial Gaussian distribution. As shown in our
results, the interaction between local scale dispersion and the
scale of the block is important in determining the time of peak
uncertainty in the dispersion tensor. The impact of the smaller
block sizes in diminishing the uncertainty in the overall dispersion
tensor is related to the fact that ergodicity is achieved, with
respect to the subgrid correlation scale, rather quickly ($L_{j}\gg
l_{j\lambda}$ for $j=2,3$ with $l_{j\lambda}$ denoting the subscale
correlation length). Furthermore, the dispersion behavior
self-averages in 3D flow fields. We also show that under finite $Pe$
conditions, lower values in the statistical anisotropy ratio of the
log-conductivity contribute to reduce the variability in the
sample-to-sample fluctuations of the large scale dispersion
behavior. This implies that dispersion self-averages faster for
lower $f_{R}$ (which is typically the conditions found in field
sites, i.e. $f_{R}\ll 1$). As expected, the initial size of the
solute plume is critical in determining the overall evolution of the
dispersion variance: For increasing transverse dimensions of the
solute plume, thus approaching ergodicity, the variance of
dispersion decreases tending to zero. Aside from the implications
associated with computation time reduction, the results shown in
this work highlights the potential of the blockscale in decreasing
the uncertainty of the effective transport behavior of the plume.
The reduction of both computational time and uncertainty in
dispersion coefficients are particularly important in the context of
probabilistic human health risk analysis where fine resolution of
the numerical grid and large number of Monte Carlo simulations are
required to accurately capture transport dynamics of multi-reactive
species and estimate the probability of extreme events
\cite{henri2015a,henri2015b}. For such reasons, blockscale
dispersion concepts could provide approximations for subscale
processes which can lead to a reduction in the computational costs.

The analysis performed in this work can be adapted to incorporate
different release conditions (in space and time), heterogeneity
correlation model and reactive transport at the cost of performing
additional numerical integrations. As an outlook to the future, a
critical issue not covered in this work is in determining the rate
in which dispersion coefficients self-average for a disordered
medium with higher levels of heterogeneity and in the presence of
sinks/sources (e.g. pumping well). Challenges remain when addressing
transport in geological formations strongly heterogeneous
permeability fields. High contrasts of the conductivity lead to
early breakthrough and late arrivals and a different upscaling
methodology might be required. For the case of high heterogeneity,
alternative upscaling approaches, such as the one presented in
Fern\`{a}ndez et al. \cite{fernandez2009b} and Li et al.
\cite{li2011} are available. It is worth mentioning that the
applicability of perturbation methods can be extended to fields with
higher heterogeneity by using the approach documented in Neuman
\cite{neuman2006}. Our analysis was limited to blocks of identical
dimensions (i.e. $\lambda_1$ = $\lambda_2$ = $\lambda_3$ =
$\lambda$). However in many applications, numerical meshes are
designed adaptively and therefore homogenized regions will have
different sizes. In this case, each numerical block would be
characterized with a distinct dispersion coefficient. The work of
Dentz and de Barros \cite{dentz_debarrosJFM15} illustrates how the
current framework can be used with a spatiotemporal \textit{dynamic}
filtering scale $\lambda_i$.

A detailed discussion on the application of blockscale dispersion
concept is provided in ch. 10, p. 253 of Rubin \cite{rubin2003book}
and Bellin et al. \cite{bellin2003IAHS}. The approach proposed by
Rubin and co-workers \cite{rubin1999,rubin2003wrr,rubin2003book}
consists of solving flow and transport over a randomly generated
coarse hydraulic conductivity field. Transport on the coarse field
is solved using the assigned average blockscale dispersion
coefficients \cite{bellin2003IAHS} which will depend on the
dimensions on the grid block. The assigned average blockscale
dispersion value is estimated from perturbation theory. In addition
to the average blockscale dispersion, we provide expressions for the
associated uncertainty epitomized by its variance. Hence, we expand
the pioneering works of Rubin and collaborators
\cite{rubin1999,rubin2003wrr} by providing expressions for the
\textit{uncertainty} in the blockscale dispersion tensor while
accounting for $Pe$, plume scale and $f_R$. The semi-analytical
expressions for the blockscale dispersion variance provided in de
Barros and Rubin \cite{deBarros2011jfm} are valid for infinite $Pe$
and $f_R$ $\ll$ 1. Our approach provides the low-order statistical
moments of the blockscale dispersion tensor and allows to randomly
assign dispersion coefficients if its probability density function
(PDF) is available. As shown in de Barros and Rubin
\cite{deBarros2011jfm}, the blockscale PDF can be obtained by
evoking entropy principles with limited information such as the
upper/lower bounds of the random variable and low order statistics,
namely the mean and variance.

The results of this work provide a quantitative basis for the use of
dispersion coefficients in coarse-grained transport models and
numerical simulations as they shed new light on the physical meaning
of different blockscale dispersion concepts and its uncertainty. The
temporal evolution of the blockscale dispersion variance provides
information relevant to the efficiency in solute mixing as well as
on ergodicity of the transport process. The uncertainty analysis of
the blockscale dispersion coefficient performed in this work
supports the conclusions of Rubin et al. \cite{rubin2003wrr} in
which plume ergodicity is achieved with respect to the blockscale.
Our findings may be relevant for the interpretation of dispersion
data from field and laboratory experiments and how they compare with
model predictions using numerical models. The concept of effective
blockscale dispersion and the associated surrogate Lagrangian and
Eulerian transport models may provide a suitable frame for the the
quantification of mixing and chemical reactions in a coarse-grained
transport framework.

\subsection*{Acknowledgements}

The first author acknowledges Yoram Rubin for fruitful discussions
on the topic of block-effective macrodispersion at the early stages
of this work. The second author acknowledges the support of the
European Research Council (ERC) through the project MHetScale
(contract number 617511). We thank all three reviewers for their
constructive comments.


\appendix

\section{Perturbation Expressions}
Using~\eqref{eq:filter},~\eqref{eq:cov-K} and~\eqref{eq:rho-FT}
in~\eqref{DensPT} and~\eqref{DeffPT}, we find that the longitudinal
ensemble and effective dispersion coefficients can be written in the
limit of $Pe \gg 1$ and $t \gg \tau_v$ as
\begin{align}
D_{11}^{ens}(t) &= D_{11} + \sqrt{\frac{\pi}{2}} \sigma_Y^2
\overline v l_1 M(1,\dots,1),
\\
D_{11}^{eff}(t) &= D_{11} + \sqrt{\frac{\pi}{2}} \sigma_Y^2
\overline v l_1 \left[M(1,\dots,1)
\right.
\nonumber\\
& \left.
- M(1,\sqrt{1+ 2 L_2^2/l_2^2 + 4 t
    /\tau_{D_2}},\dots,\sqrt{1 + 2 L_d^2/l_d^2 + 4 t
  /\tau_{D_d}}) \right],
\end{align}
where the auxiliary function $M(\mathbf A)$ is given by

\begin{align}
&M(\mathbf A) = \sqrt{\frac{2}{\pi}}\int\limits_0^{\infty} dt^\prime \left[\int_k
  \exp\left(-\frac{k_i^2A_i^2}{2} -ik_1 t^\prime\right) -
\right.
\nonumber\\
& \left.
\prod_{i=1}^d
  \int\limits_{-\pi/\lambda_i}^{\pi/\lambda_i} \frac{d k_i}{2\pi}
  \exp\left(-\frac{k_i^2A_i^2}{2} -ik_1 t^\prime\right) \right] +
\dots
\end{align}
where the dots denote terms of order $Pe^{-1}$ and $\tau_v/t$.
We note that
\begin{align}
\int\limits_0^\infty d t^\prime \exp(-ik_1t^\prime) = \pi \delta(k_1).
\end{align}
Thus, we obtain for $M(\mathbf A)$
\begin{align}
M(\mathbf A) = M(1,A_2,\dots,A_d) = \frac{1 - \prod_{i=2}^d {\rm
    erf}\left(\frac{\pi}{\sqrt{2} \lambda_i \sqrt{A_i}} \right)}{\prod_{i=2}^d \sqrt{A_i}}.
\end{align}

\bibliographystyle{elsarticle-num}

\end{document}